\documentclass[preprint,3p,times,twocolumn]{elsarticle}

\usepackage{amsmath,amssymb,amsfonts,amsthm,enumerate,multirow,mathtools}
\usepackage{color}
\usepackage{siunitx}
\usepackage[linesnumbered,ruled]{algorithm2e} 
\usepackage{diagbox} 
\usepackage{cleveref}
\usepackage{placeins} 
\usepackage[table]{xcolor}
\usepackage{booktabs}
\usepackage{natbib}
\usepackage{placeins}
\usepackage{graphicx}

\newcommand{\redchanges}[1]{{\color{black}#1}}

\newcommand{\pmat}[1]{\begin{bmatrix}#1\end{bmatrix}}
\DeclareMathOperator*{\argmax}{argmax}
\newcommand{\Transp}{\mathsf{T}}

\usepackage{lineno}

\journal{ArXiv}

\begin{document}

\begin{frontmatter}
\newcommand{\coverTitle}{A Bayesian Approach to Triaxial Strain Tomography from High-energy X-ray Diffraction}
\newcommand{\coverAuthors}{J.N. Hendriks, C.M. Wensrich, A. Wills.}
\newcommand{\coverStatus}{Accepted for publication.}

\begin{titlepage}
    \begin{center}
        {\large \em Technical report}
        
        \vspace*{2.5cm}
        %
        {\Huge \bfseries \coverTitle  \\[0.4cm]}
        
        %
        {\Large \coverAuthors \\[2cm]}
        
        \renewcommand\labelitemi{\color{red}\large$\bullet$}
        \begin{itemize}
            \item {\Large \textbf{Please cite this version:}} \\[0.4cm]
            \large
            \coverAuthors. \coverTitle. \textit{Strain. 2020:e12341. doi: 10.1111/str.12341}  
        \end{itemize}
        
        \vfill

        
        \vfill
        \vspace{50mm}
    \end{center}
\end{titlepage}
\newpage
\thispagestyle{empty}
\newpage

\title{A Bayesian Approach to Triaxial Strain Tomography from High-energy X-ray Diffraction}

\author[NewcE]{J.N. Hendriks}
\author[NewcE]{C.M. Wensrich}
\author[NewcE]{A. Wills}

\address[NewcE]{School of Engineering, The University of Newcastle, Callaghan NSW 2308, Australia}

\begin{abstract}
Diffraction of high-energy X-rays produced at synchrotron sources can provide rapid strain measurements, with high spatial resolution, and good penetrating power.
With an uncollimated diffracted beam, through thickness averages of strain can be measured using this technique, which poses an associated rich tomography problem.
This paper proposes a Gaussian process (GP) model for three-dimensional strain fields satisfying static equilibrium and an accompanying algorithm for tomographic reconstruction of strain fields from high-energy X-ray diffraction.
\redchanges{We present numerical evidence that} this method can achieve triaxial strain tomography in three-dimensions using only a single axis of rotation.
The method builds upon recent work where the GP approach was used to reconstruct two-dimensional strain fields from neutron based measurements.
A demonstration is provided from simulated data, showing the method is capable of rejecting realistic levels of Gaussian noise.
\end{abstract}


\end{frontmatter}

\section{Introduction} 
\label{sec:introdution}
Diffraction of X-rays and neutrons allow the study of mechanical stress and strain within crystalline solids \cite{kisi08,noyan87,hauk97,fitzpatrick03}.
These techniques revolve around Bragg's law; $\lambda = 2d\sin\theta$, whereby variations in the lattice spacing $d$, due to elastic strain can be observed by changes in the scattering angle, $\theta$, of the diffracted radiation with wavelength $\lambda$. 
These variations are related to the average elastic strain within the scattering volume according to
\begin{equation}\label{eq:measured_strain}
    \langle \epsilon \rangle = \frac{d - d_0}{d_0},
\end{equation}
where $d_0$ is the lattice spacing in a strain free sample.
Both neutron and  conventional X-ray diffraction have limitations; lab-based X-rays sample only a shallow surface layer (typically a few microns), while the characteristic low intensities of neutrons gives rise to long acquisition times and spatial resolutions of $1\text{mm}$ or larger.

Modern X-ray synchrotron sources can provide very intense narrow beams of highly penetrating X-ray photons \cite{withers2002residual}.
These high-energy X-rays can provide strain measurements with beam spot sizes as small as $1\mu\text{m}$ and can have path lengths of many centimetres, even in steel. Although high-energy X-ray diffraction can be used to study dynamic strains (e.g. \cite{mostafavi2017dynamic}), in this paper the problem is restricted to steady state. The restriction to steady-state allows the static equilibrium constraints to be assumed as prior information.

A particular feature of high energy X-rays is that the scattering angle is typically small $2\theta < 15^\circ$ \cite{withers2002residual}.
This has two implications for the study of strain fields: firstly, that the normal component of strain measured is almost perpendicular to the incident beam; at an angle of $90^\circ - \theta$; secondly, the gauge volume is elongated in the direction of the incident beam (typically by a factor of 10 or more).

The ability to select small beam cross-sections allows strain profiles \cite{croft2005strain} or two dimensional strain maps \cite{withers2002residual} to be readily obtained in the plane perpendicular to the incident beam (see Figure~\ref{fig:x_ray_setup} for the measurement geometry). 
However, the resolution along the incident beam is degraded by a factor of 10 or more due to the elongation of the scattering volume.
This does not always present a problem, for example in \cite{croft2005strain} the through thickness strain variation in the direction of the incident beam was known to be small, and averaging in this direction provided a good measure of of strain.

To overcome these limitations, a different approach is presented in \cite{korsunsky2006principle} and \cite{korsunsky2011strain} where determining the strain field is considered a rich tomography problem from a series of through thickness measurements. 
In this setting, the goal is to reconstruct the higher order (two- or three-dimensional) distribution of unknown strain from a set of lower dimension (one- or two-dimensional) projections.
In \cite{korsunsky2006principle}, the axisymmetric strain within a quenched cylinder was reconstructed, providing an initial demonstration of this conceptual approach.
This approach was extended in \cite{korsunsky2011strain}, where reconstruction of the longitudinal strain within a zirconia dental prostheses was achieved by posing the problem in a form suitable for conventional computed tomography algorithms. 
In this experiment, the sample was rotated about a single axis while through-thickness averages of the out-of-plane normal strain were recorded.
The resulting scalar tomography problem was then solved using conventional back-projection techniques.

This problem is also studied by Lionheart and Withers\cite{lionheart15}, where the transverse ray transform \cite{sharafutdinov1994integral} is given as a possible model for the high-energy X-ray strain measurements \cite{lionheart15}. This assumes that the direction of measured strain is exactly perpendicular to the beam, i.e. that the angle of diffraction is close to zero.
Further, It is proposed that full triaxial strain field tomography can be done as a series of `regular' tomography problems.
Here, each regular tomography involves rotating the sample about a single axis and reconstructing the component of strain in the direction of the axis of rotation.
By performing six of these experiments about different axes of rotation the full triaxial strain field could be determined. 
The conditions on the the axes chosen to allow for the the triaxial strain to be reconstructed are given.

This work is extended in Desai and Lionheart\cite{Desai_2016} through the presentation of an explicit inversion formula for the transverse ray transform. In general, it was shown that this formula allows for the reconstruction of strain from high-energy X-ray measurements made around three axes of rotation. It was also shown that if compatibility can be assumed then only two axes of rotation are required. Decreasing the number of axes that the sample has to be rotated about reduces the complexity of the experiment set up and the time required to run the experiment.

In this paper, a Gaussian process based approach to reconstruct the triaxial strain field from these measurements is presented.
This approach enforces that the reconstruction to satisfies the static equilibrium constraints, and is applied to the problem of reconstructing the strain field using single axis tomography with promising results in simulation.
Recent work in related fields has demonstrated the use of Gaussian processes (GPs) to model and reconstruct two-dimensional strain fields from time-of-flight neutron transmission measurements \cite{jidling2018probabilistic,hendriks2018traction} and neutron diffraction measurements \cite{hendriks2018robust}.

\subsection{Contribution} 
\label{sub:contribution}
This paper makes the following contributions:
\begin{enumerate}
    \item A Gaussian process model for steady-state, triaxial elastic strain fields in three-dimensions. This is a non-trivial extension of the model presented in Jidling et al \cite{jidling2018probabilistic} which was restricted to biaxial strain fields in two dimensions under the assumption of plane stress or plane strain.
    \item A method for applying this model to reconstruct the strain field from high-energy X-ray strain measurements. Results indicate \redchanges{that by assuming equilibrium} this approach can reconstruct a triaxial strain field using a single axis of rotation. This could not be achieved using existing methods.
\end{enumerate}


\section{High Energy X-ray Strain Measurement} 
\label{sec:x_ray_strain_measurement}
This section provides a brief summary of strain measurement using high energy X-ray diffraction. 
This process is not the primary focus of this work, however the summary provides details pertinent to model the relationship between the measurements and the strain field.
A more detailed description can be found in \cite{withers2002residual} and details of geometric corrections required due to rotating the sample causing changes in the sample to detector distance are describe in \cite{korsunsky2011strain}.
The summary is as follows.

\begin{figure}[!ht]
    \centering
    \includegraphics[width=1.0\linewidth]{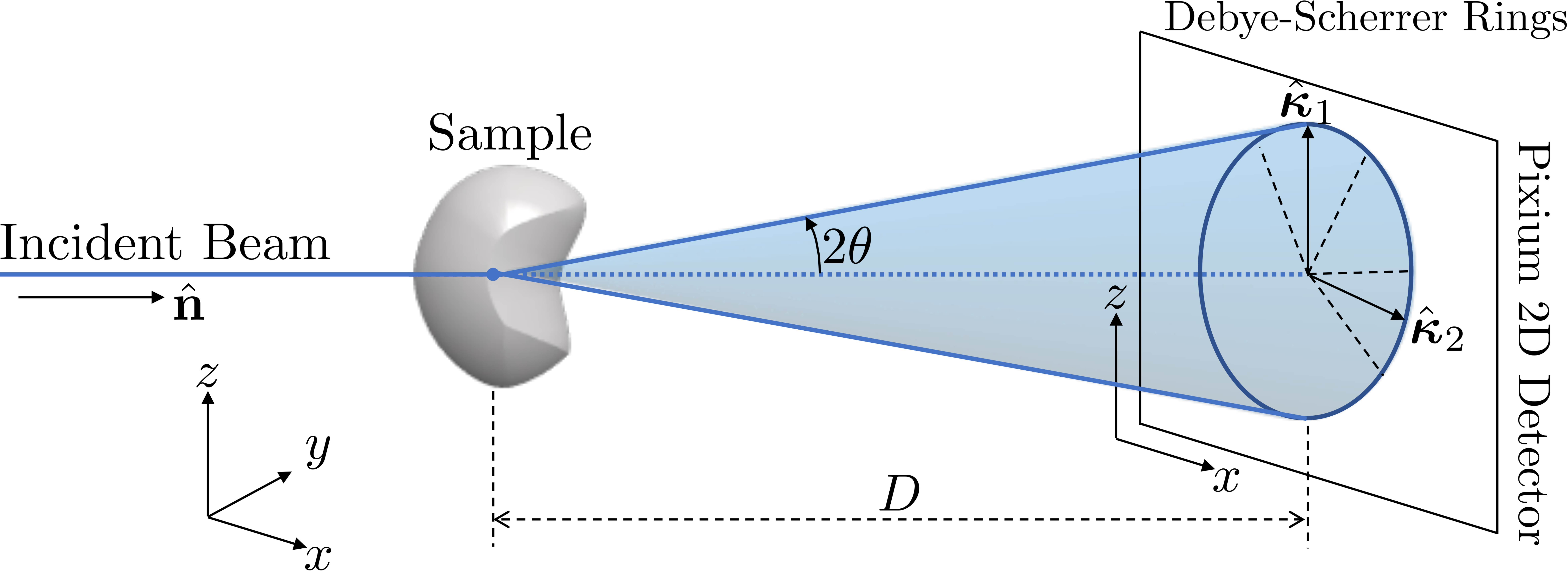}
    \caption{Measurement geometry. The incident beam, with direction $\hat{\mathbf{n}}$, is diffracted by the sample at an angle of $2\theta$ forming a shallow cone. The peak intensities of this cone, known as a Debye-Scherrer ring, are recorded by a detector at distance $D$.
    Two possible strain measurement directions, $\hat{\boldsymbol\kappa}_1$ and $\hat{\boldsymbol\kappa}_2$, are shown, which related to a measurement of strain given by analysing their corresponding segments of the Debye-Scherrer ring (indicated by the dashed lines). Figure is not to scale. }
    \label{fig:x_ray_setup}
\end{figure}

The incident beam with direction $\hat{\mathbf{n}}$ is diffracted within the sample according to Bragg's law at an angle of $2\theta$ forming a shallow cone. 
The intensity of the diffracted beam is recorded at a detector, with the peak intensity's forming a Debye-Scherrer ring, see Figure~\ref{fig:x_ray_setup}. 
Due to the small diffraction angle, the distance to the detector, $D$, is much greater than nominal sample dimensions.

For a polychromatic X-ray beam, a diffraction pattern can be fit to a segment of the Debye-Scherrer ring providing a measurement of the average normal strain in the direction$\kappa$ of the form \eqref{eq:measured_strain}; with a $10^\circ$ being suitable for strain measurements \cite{korsunsky2011strain}.
Two possible segments and are shown in Figure~\ref{fig:x_ray_setup} with corresponding measurement directions $\hat{\boldsymbol\kappa}_1$ and $\hat{\boldsymbol\kappa}_2$, respectively.
Multiple segments can be analysed from each Debye-Scherrer ring to give measurements of the normal strain in different directions. 
Although using the relative shift of the diffraction pattern has been done in practice, Lionheart and Withers\cite{lionheart15} shows that, theoretically, a particular moment of the diffraction pattern should be used instead.

\begin{figure}[!ht]
    \centering
    \includegraphics[width=0.7\linewidth]{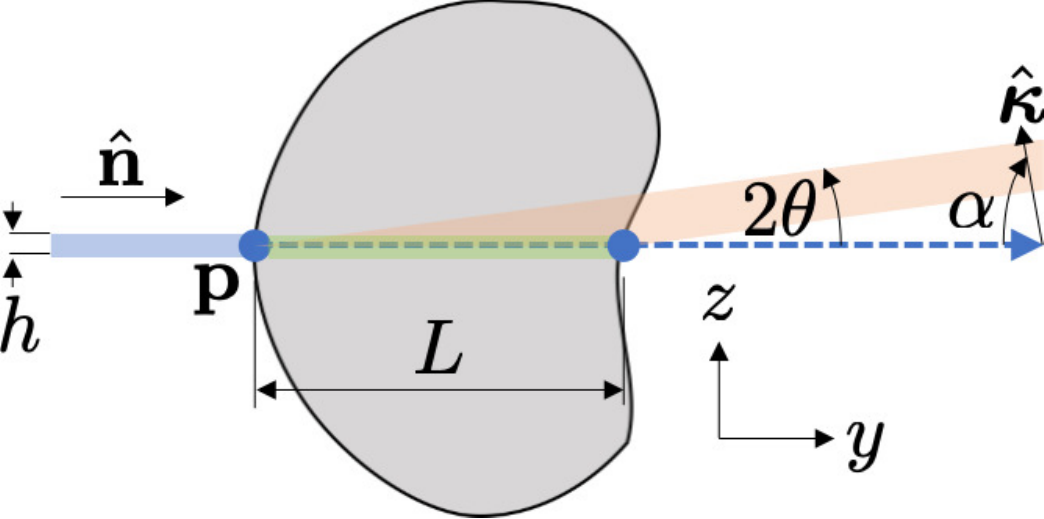}
    \caption{
    Geometry of the scattering volume. Shown is a cross section of the sample, the incident (blue) and the diffracted (orange) beams, and the scattering volume (green) in the plane defined by $\hat{\mathbf{n}}$ and $\hat{\boldsymbol\kappa}$. The incident beam spot size $h\times h$, $L$ is the through thickness length of the scattering volume, $\alpha = 90-\theta$ is the angle from the incident beam to the measured normal strain direction, and $\mathbf{p}$ is the initial intersection of the beam and the sample. Only the top half of the diffracted beam is shown, for clarity. Figure not to scale.
    }
    \label{fig:LRT_geom}
\end{figure}

In this work, we consider the diffracted beam to be left uncollimated---which aligns with the work in \cite{korsunsky2011strain}, and the incident beam to be collimated to give a spot size of $h\times h$.
This measurement geometry is shown in Figure~\ref{fig:LRT_geom} and gives the scattering volume's length as the path length through the sample, $L$.

A measurement model can be formulated using the following reasoning. The measurement corresponds to an average of normal strain in direction $\hat{\boldsymbol\kappa}$ over the gauge volume. This average can be represented by a volume integral divided by the volume. However, since spot sizes as small as $1\mu\text{m}$ are achievable and that typical path lengths can be several orders of magnitude larger, it is reasonable to assume that there is no variation in the perpendicular to the direction of the beam. 
Under this assumption, the volume integral can be reduced to a line integral in the beam direction;
\begin{equation}
\begin{split}\label{eq:sTRT}
    I_\epsilon &= \frac{1}{L}\int\limits_0^L \hat{\boldsymbol\kappa}^\Transp\boldsymbol\epsilon(\mathbf{p}+\hat{\mathbf{n}}s)\hat{\boldsymbol\kappa}\,\mathrm{d}s + e, \\
\end{split}
\end{equation}
where $e\sim\mathcal{N}(0,\sigma_n^2)$, $\boldsymbol\epsilon(\mathbf{x})$ is the strain tensor field inside the sample and $\mathbf{p}$ is the initial intersection between the beam and the sample.

By analysing multiple segments from each Debye-Scherrer ring, each ray provides information about multiple components of strain. For example, assuming that the measurement direction is exactly perpendicular to the beam direction. Then, choosing at least three unique measurement directions, each ray provides information about the strain projected onto the plane perpendicular to the beam. That is, a beam with direction $\hat{\mathbf{n}} = [1,0,0]^{\Transp}$ predominantly provides information about $\epsilon_{yy}$, $\epsilon_{yz}$, and $\epsilon_{zz}$.

If we make the additional assumption that the measurement directions are perpendicular to the incident beam, then this line of reasoning also leads to the transverse ray transform (TRT) which was proposed by \cite{lionheart15} to model high-energy X-ray strain measurements. With the addition of noise, the TRT model for the measurement is given by
\begin{equation}
     J_\epsilon = \frac{1}{L}\int_{0}^{L} \Pi_{\hat{\mathbf{n}}}\epsilon(\mathbf{p}+\hat{\mathbf{n}}s)\Pi_{\hat{\mathbf{n}}},
 \end{equation} 
where 
\begin{equation}
    \Pi_{\hat{\mathbf{n}}} = \mathbf{I} - \hat{\mathbf{n}}\hat{\mathbf{n}}^{\Transp},
\end{equation}
and $\mathbf{I}$ is the identity matrix. Here, $\Pi_{\hat{\mathbf{n}}}\epsilon\Pi_{\hat{\mathbf{n}}}$ is the projection of the strain field onto the plane perpendicular to $\hat{\mathbf{n}}$. For example, consider that $\hat{\mathbf{n}} = [1, 0, 0]^{\Transp}$, then
\begin{equation}
    \Pi_{\hat{\mathbf{n}}}\epsilon\Pi_{\hat{\mathbf{n}}} = 
    \pmat{0 & 0 & 0 \\
          0 & \epsilon_{yy} & \epsilon_{yz} \\
          0 & \epsilon_{yz} & \epsilon_{zz}}.
\end{equation}
Although the TRT represents a more compact measurement model, it requires assuming the measurement directions are perpendicular to the beam direction, which is not quite the case in practice. Hence, \eqref{eq:sTRT} has been used as the measurement model in this work.


\section{3D Strain field Reconstruction} 
\label{sec:strain_field_reconstruction}
In this work, the strain field is reconstructed using the framework of Bayesian inference. 
The strain field is modelled as having a probability distribution, this does not mean that the strain field is random, instead it represents our uncertainty in its values.
A \emph{prior} probability distribution is assigned to the strain field, $p(\boldsymbol\epsilon)$, that represents any knowledge we have before the inclusion of measurements, $y$.
This distribution is then updated by the inclusion of measurements, which are assigned a \emph{likelihood} $p(y | \boldsymbol\epsilon)$, using Bayes' rule to give a \emph{posterior} distribution of the strain:
\begin{equation}
    p(\boldsymbol\epsilon | y) = \frac{p(y | \boldsymbol\epsilon)p(\boldsymbol\epsilon)}{p(y)}.
\end{equation}

As the strain field must satisfy equilibrium, the prior distribution should incorporate this knowledge. 
In two dimensions this can be achieved by modelling the Airy's stress function by a Gaussian process (GP) from which the two-dimensional strain field can be defined under an assumption of plane stress or plane strain \cite{jidling2018probabilistic}.
This method was successfully used to reconstruct two-dimensional strain fields from neutron transmission measurements \cite{jidling2018probabilistic,hendriks2018traction} and neutron diffraction measurements \cite{hendriks2018robust}.

The following sections present a generalisation of this procedure to three-dimensions and application to reconstructing the strain field from from high energy X-ray diffraction measurements. 
A brief overview of GPs is provided in Section~\ref{sec:gaussian_processs}. 
A Gaussian process prior suitable for modelling three-dimensional strain fields is designed in Section~\ref{sec:modelling_three_dimensional_strain_fields}.
A likelihood model for high energy X-ray strain measurements is then defined in Section~\ref{sec:xray_measurement_model}, allowing the strain field to be reconstructed.
For this work, the material should have a randomly distributed polycrystalline structure with no preferred crystal orientation (i.e. no texture) such that the bulk material is isotropic.

\subsection{Gaussian Process} 
\label{sec:gaussian_processs}
This section gives a brief description of GPs, for a more detailed description see Rasmussen and Williams\cite{rasmussen2006gaussian}. 
A GP is a stochastic process suitable for modelling spatially correlated functions and can be viewed as a distribution over functions;
\begin{equation}
    f(\mathbf{x})\sim\mathcal{GP}\left(m(\mathbf{x}),k(\mathbf{x},\mathbf{x}')\right),
\end{equation}
where $m(\mathbf{x}) = \mathbb{E}\left[f(\mathbf{x})\right]$ is the mean function and $k(\mathbf{x},\mathbf{x}')  = \mathbb{E}\left[\left(m(\mathbf{x}) - f(\mathbf{x})\right)\left(m(\mathbf{x}) - f(\mathbf{x})\right)^\Transp\right]$ is the covariance function.
The choice of covariance function governs the characteristics of the functions in this distribution---such as their smoothness. 
Many choices exist for the covariance function and a good summary is available elsewhere \cite{rasmussen2006gaussian}.

A Gaussian process is a generalisation of a multivariate Gaussian in the sense that function values sampled at a finite number of inputs $\mathbf{x}_1,\dots,\mathbf{x}_N$ are Gaussian distributed;
\begin{subequations}
\begin{equation}
    \begin{bmatrix}
        f(\mathbf{x}_1) \\\vdots \\f(\mathbf{x}_N) 
    \end{bmatrix}\sim \mathcal{N}\left(\boldsymbol{\mu},\mathbf{K}\right) \quad \text{where} \quad \mu = \begin{bmatrix}
        m(\mathbf{x}_1) \\ \vdots \\ m(\mathbf{x}_N)
    \end{bmatrix},
\end{equation}
and
\begin{equation}
    \mathbf{K} = \begin{bmatrix}
        k(\mathbf{x}_1,\mathbf{x}_1) & \cdots & k(\mathbf{x}_1,\mathbf{x}_N) \\
        \vdots & \ddots & \vdots \\
        k(\mathbf{x}_N,\mathbf{x}_1) & \cdots & k(\mathbf{x}_N,\mathbf{x}_N)
    \end{bmatrix}.
\end{equation}
\end{subequations}

If a GP prior is used and the measurement likelihood is Gaussian then posterior distribution given by Bayes' rule has a closed form and can be computed using standard Gaussian conditioning.


\subsection{Three-dimensional Strain Field GP} 
\label{sec:modelling_three_dimensional_strain_fields}
Here, we design a GP prior for strain in three-dimensions that intrinsically satisfies static equilibrium. 
The strain field is defined as the symmetric tensor
\begin{equation}
    \boldsymbol\epsilon(\mathbf{x}) = \begin{bmatrix}
        \epsilon_{xx}(\mathbf{x}) & \epsilon_{xy}(\mathbf{x}) & \epsilon_{xz}(\mathbf{x}) \\
        \epsilon_{xy}(\mathbf{x}) & \epsilon_{yy}(\mathbf{x}) & \epsilon_{yz}(\mathbf{x}) \\
        \epsilon_{xz}(\mathbf{x}) & \epsilon_{yz}(\mathbf{x}) & \epsilon_{zz}(\mathbf{x})
    \end{bmatrix},
\end{equation}
where $\mathbf{x} = [x \ \ y \ \ z]^\Transp$ are the spatial coordinates.

This GP will be of the form
\begin{equation}
    \bar{\boldsymbol\epsilon}\sim\mathcal{GP}\left(\mathbf{0},\mathbf{K}_{\epsilon}(\mathbf{x},\mathbf{x}')\right),
\end{equation}
where $\bar{\boldsymbol\epsilon}$ is a vector of the $6$ unique components of the strain field.
Here, the covariance function for the strain field, $\mathbf{K}_{\epsilon}(\mathbf{x},\mathbf{x}')$ will be designed to ensure that all strain fields belonging to this GP satisfy equilibrium.
This is done by specifying a GP prior for a set of potentials known as the Beltrami stress functions.
Having done this, a GP prior for the strain field can be derived using the equilibrium equations and Hooke's law.
The details of this derivation are as follows.

The Beltrami stress functions \cite{beltrami1892osservazioni} allow a complete solution to the equilibrium equations in three-dimensions to be written as \cite{szeidl1996complete,sadd2009elasticity,carlson1966completeness};
\begin{equation}
 \boldsymbol\sigma(\mathbf{x}) = \boldsymbol\nabla \times \boldsymbol\Phi(\mathbf{x}) \times \boldsymbol\nabla,
\end{equation}
where $\boldsymbol\nabla = \pmat{\frac{\partial}{\partial x} & \frac{\partial}{\partial y} & \frac{\partial}{\partial z}}^\Transp$ and $\boldsymbol\sigma(x,y,z)$ is the symmetric stress tensor;
\begin{equation}
    \boldsymbol\sigma(x,y,z) = \begin{bmatrix}
        \sigma_{xx}(\mathbf{x}) & \sigma_{xy}(\mathbf{x}) & \sigma_{xz}(\mathbf{x}) \\
        \sigma_{xy}(\mathbf{x}) & \sigma_{yy}(\mathbf{x}) & \sigma_{yz}(\mathbf{x}) \\
        \sigma_{xz}(\mathbf{x}) & \sigma_{yz}(\mathbf{x}) & \sigma_{zz}(\mathbf{x})
    \end{bmatrix},
\end{equation}
and $\boldsymbol\Phi(x,y,z)$ is the Beltrami stress tensor consisting of six unique scalar potential fields;
\begin{equation}
    \boldsymbol\Phi(\mathbf{x}) = \begin{bmatrix}
        \Phi_{1}(\mathbf{x}) & \Phi_{4}(\mathbf{x}) & \Phi_{5}(\mathbf{x}) \\
        \Phi_{4}(\mathbf{x}) & \Phi_{2}(\mathbf{x}) & \Phi_{6}(\mathbf{x}) \\
        \Phi_{5}(\mathbf{x}) & \Phi_{6}(\mathbf{x}) & \Phi_{3}(\mathbf{x}) \\
    \end{bmatrix}.
\end{equation}

To improve readability we introduce the following vectorised notation;
\begin{equation}
    \bar{\boldsymbol\Phi} = \begin{bmatrix}\Phi_{1} \\ \Phi_{2} \\ \Phi_{3} \\ \Phi_{4} \\ \Phi_{5} \\ \Phi_{6}\end{bmatrix}, \quad 
    \bar{\boldsymbol\sigma} = \begin{bmatrix}
        \sigma_{xx} \\ \sigma_{yy} \\ \sigma_{zz} \\
        \sigma_{xy} \\ \sigma_{xz} \\ \sigma_{zz} \\
    \end{bmatrix}, \quad 
        \bar{\boldsymbol\epsilon} = \begin{bmatrix}
        \epsilon_{xx} \\ \epsilon_{yy} \\ \epsilon_{zz} \\
        \epsilon_{xy} \\ \epsilon_{xz} \\ \epsilon_{yz} \\
    \end{bmatrix},
\end{equation}
where the spatial coordinates $\mathbf{x}$, are omitted for brevity. 
Using this notation we can write Hooke's law which relates the stress and strain fields as 
\begin{equation}
\begingroup
\setlength\arraycolsep{1.8pt}
    \begin{bmatrix}
        \epsilon_{xx} \\ \epsilon_{yy} \\ \epsilon_{zz} \\
        \epsilon_{xy} \\ \epsilon_{xz} \\ \epsilon_{yz} \\
    \end{bmatrix} = 
    \frac{1}{E}\underbrace{\begin{bmatrix}
        -\nu & 1 & -\nu & 0 & 0 & 0 \\
        -\nu & -\nu & 1 & 0 & 0 & 0 \\
        0 & 0 & 0 & 1+\nu & 0 & 0 \\ 
        0 & 0 & 0 & 0 & 1+\nu & 0 \\
        0 & 0 & 0 & 0 & 0 & 1+\nu \\
    \end{bmatrix}}_{\mathbf{H}}
    \begin{bmatrix}
        \sigma_{xx} \\ \sigma_{yy} \\ \sigma_{zz} \\
        \sigma_{xy} \\ \sigma_{xz} \\ \sigma_{zz} \\
    \end{bmatrix},\endgroup
\end{equation}
where $\mathbf{H}$ is the compliance matrix for isotropic materials, $\nu$ is Poisson's ratio, and $E$ is Young's modulus. The scaling $\frac{1}{E}$ can be safely neglected in the methods implementation to provide better numerical scaling.
Finally, we can write the mapping from the Beltrami stress functions to the strain field in this vectorised form as
\begin{equation}
    \bar{\boldsymbol\epsilon} = \mathbf{H}\mathcal{B}^\mathbf{x}\bar{\boldsymbol\Phi},
\end{equation}\label{eq:beltrami_to_strain}
where

\begin{equation}
\begingroup
\setlength\arraycolsep{1.7pt}
    \mathcal{B}^\mathbf{x} = \begin{bmatrix}
        0 & \frac{\partial^2}{\partial z^2} & \frac{\partial^2}{\partial y^2} & 0 & 0 & -2\frac{\partial^2}{\partial y \partial z} \\
        \frac{\partial^2}{\partial z^2} & 0 & \frac{\partial^2}{\partial x^2} & 0 & -2\frac{\partial^2}{\partial x \partial z} & 0 \\
        \frac{\partial^2}{\partial y^2} & \frac{\partial^2}{\partial x^2} & 0 & -2\frac{\partial^2}{\partial x \partial y} & 0 & 0 \\
        0 & 0 & -\frac{\partial^2}{\partial x \partial y} & -\frac{\partial^2}{\partial z^2} & \frac{\partial^2}{\partial y \partial z} & \frac{\partial^2}{\partial x \partial z} \\
        -\frac{\partial^2}{\partial y \partial z} & 0 & 0 & \frac{\partial^2}{\partial x \partial z} & \frac{\partial^2}{\partial x \partial y} & -\frac{\partial^2}{\partial x^2} \\
        0 & -\frac{\partial^2}{\partial x \partial z} & 0 & \frac{\partial^2}{\partial y \partial z} & -\frac{\partial^2}{\partial y^2} & \frac{\partial^2}{\partial x \partial y},
    \end{bmatrix}\endgroup
\end{equation}
and the superscript is used to denote the set of spatial coordinates on which the operator acts.

As this mapping is linear it can be used to define a GP prior on the strain function \cite{wahlstrom2015modeling,rasmussen2006gaussian,jidling2018probabilistic}. 
Each component of the Beltrami tensor field, $\Phi_i(\mathbf{x})$, is assigned a Gaussian process prior with its own covariance function $k_i(\mathbf{x},\mathbf{x}')$ --- the squared-exponential is used in Section~\ref{sec:simulation}. The covariance function encodes spatial correlation which assumes the resulting strain field will have a degree of smoothness.
Each component of the Beltrami tensor is considered independent as their is no prior information to suggest otherwise, and including correlation between the components would result in the strain field being restricted to meet an additional conditions beyond the desired equilibrium condition.
This gives a GP for $\bar{\boldsymbol\Phi}(\mathbf{x})$ as
\begin{equation}
\begingroup
\setlength\arraycolsep{3.5pt}
    \bar{\boldsymbol\Phi}(\mathbf{x}) \sim \mathcal{GP}\left(
    \underbrace{\begin{bmatrix}
        m_{1} \\ m_{2} \\ m_{3} \\
        m_{4} \\ m_{5} \\ m_{6}
    \end{bmatrix}}_{\mathbf{m}_\Phi(\mathbf{x})},
    \underbrace{\begin{bmatrix}
        k_{1} & 0 & 0 & 0 & 0 & 0 \\
        0 & k_{2} & 0 & 0 & 0 & 0 \\
        0 & 0 & k_{3} & 0 & 0 & 0 \\
        0 & 0 & 0 & k_{4} & 0 & 0 \\
        0 & 0 & 0 & 0 & k_{5} & 0 \\
        0 & 0 & 0 & 0 & 0 & k_{6}\\
    \end{bmatrix}}_{\mathbf{K}_\Phi(\mathbf{x},\mathbf{x}')}\right),
    \endgroup
\end{equation}
where the shorthand $k_i = k_i(\mathbf{x},\mathbf{x}')$ and $m_i = m_i(\mathbf{x})$ has been used.

The mapping~\eqref{eq:beltrami_to_strain} is applied to give a GP prior for the strain field that will ensure that any estimated strain field satisfies equilibrium;
\begin{equation}
\begin{split}
    \bar{\boldsymbol\epsilon}&\sim\mathcal{GP}\left(\mathbf{H}\mathcal{B}^\mathbf{x}\mathbf{m}_\Phi,\mathbf{H}\mathcal{B}^\mathbf{x}\mathbf{K}_\Phi(\mathbf{x},\mathbf{x}')\mathcal{B}^{\mathbf{x}'}{}^\Transp\mathbf{H}^\Transp\right) \\
    &= \mathcal{GP}\left(\mathbf{0},\mathbf{K}_{\epsilon}(\mathbf{x},\mathbf{x}')\right).
\end{split}
\end{equation}
The covariance function for the strain field, $\mathbf{K}_\epsilon$, has correlations between the individual components of strain that ensure any estimated strain field will satisfy equilibrium.
Here, without loss of generality \cite{rasmussen2006gaussian}, we chose the prior mean functions as zero.


\subsection{Reconstruction from X-ray Strain Measurements} 
\label{sec:xray_measurement_model}
Here, we define the likelihood of the X-ray strain measurements $p(I_\epsilon | \boldsymbol\epsilon)$. 
In vector form we can write the measurement model as 
\begin{equation}\label{eq:measurement_operator}
\begin{split}
    I_\epsilon(\boldsymbol\eta) &= \frac{1}{L}\int\limits_0^L \bar{\boldsymbol\kappa}\bar{\boldsymbol\epsilon}(\mathbf{p}+\hat{\mathbf{n}}s)\,\mathrm{d}s + e \\
    I_\epsilon(\boldsymbol\eta) &= \mathcal{L}^\mathbf{x}(\boldsymbol{\eta})\bar{\boldsymbol\epsilon}(\mathbf{x}) + e
\end{split}
\end{equation}
where $\bar{\boldsymbol\kappa} = \pmat{\kappa_x^2 & \kappa_y^2 & \kappa_z^2 & 2\kappa_x\kappa_y & 2\kappa_x\kappa_z & 2\kappa_y\kappa_z}$, $\boldsymbol\eta = \left\{\boldsymbol\kappa,\hat{\mathbf{n}},\mathbf{p},L\right\}$, $e\sim\mathcal{N}(0,\sigma_n^2)$, and $\mathcal{L}^\mathbf{x}(\boldsymbol{\eta})$ is a considered an operator that maps from the strain function $\bar{\boldsymbol\epsilon}(\mathbf{x})$ into the measurements $I_\epsilon(\boldsymbol\eta)$.
As this operator is linear, the joint distribution of the strain field at user specified location of interest, $\bar{\boldsymbol\epsilon}_* = \bar{\boldsymbol\epsilon}(\mathbf{x}_*)$, and the measurements $\mathbf{I}_\epsilon = \pmat{I_\epsilon(\boldsymbol\eta_1)&\cdots&I_\epsilon(\boldsymbol\eta_n)}^\Transp$ is Gaussian \cite{wahlstrom2015modeling,rasmussen2006gaussian,jidling2018probabilistic};

\begin{equation}\label{eq:joint_distribution}
    \begin{bmatrix}
        \mathbf{I}_\epsilon \\
        \bar{\boldsymbol\epsilon}_*
    \end{bmatrix}
     = \mathcal{N}\left(\begin{bmatrix}
         \mathbf{0} \\ \mathbf{0}
     \end{bmatrix},\begin{bmatrix}
         \mathbf{K}_{I}+\sigma_n^2\mathbf{I} & \mathbf{K}_{*}^\Transp \\
         \mathbf{K}_{*} & \mathbf{K}_{\epsilon}
     \end{bmatrix}\right),
\end{equation}
where $\mathbf{K}_*$ is the cross covariance between the strains and the measurements and $\mathbf{K}_I$ is the covariance of the measurements. 

The cross covariance between the strain $\bar{\boldsymbol\epsilon}_*$ and a measurement $I_\epsilon(\boldsymbol\eta_i)$ is given by a single application of \eqref{eq:measurement_operator} to the strain fields covariance function;
\begin{equation}
    \begin{split}
        (\mathbf{K}_{*})_{i} &= \mathbf{K}_{\epsilon}\mathcal{L}^{\mathbf{x}'}{}(\boldsymbol\eta_i)^\Transp \\ &= \frac{1}{L_i}\int\limits_0^{L_i}\mathbf{K}_\epsilon(\mathbf{x}_*,\mathbf{p}_{i}+\hat{\mathbf{n}}_is')\bar{\boldsymbol\kappa}_i^\Transp\,\mathrm{d}s'
    \end{split}
\end{equation}
and the covariance between each pair of measurements, $I_\epsilon(\boldsymbol\eta_i)$ and $I_\epsilon(\boldsymbol\eta_j)$ is similarly given by
\begin{equation}
\begin{split}
    (\mathbf{K}_I)_{ij} &= \mathcal{L}^\mathbf{x}(\boldsymbol\eta_i)\mathbf{K}_\epsilon\mathcal{L}^{\mathbf{x}'}(\boldsymbol\eta_j)^\Transp \\
    &= \frac{1}{L_iL_j}\int\limits_0^{L_i}\hspace{-2mm}\int\limits_0^{L_j}\bar{\boldsymbol\kappa}_i\mathbf{K}_\epsilon(\mathbf{p}_i+\hat{\mathbf{n}}_is,\mathbf{p}_j+\hat{\mathbf{n}}_js')\bar{\boldsymbol\kappa}_j^\Transp \,\mathrm{d}s'\mathrm{d}s
\end{split}
\end{equation}

The posterior distribution of the strain $\bar{\boldsymbol\epsilon}$ conditioned on the measurements $\mathbf{I}_\epsilon$ is
\begin{equation}
    \bar{\boldsymbol\epsilon}_* \sim \mathcal{N}(\boldsymbol\mu_{\bar{\epsilon}|\mathbf{I}_\epsilon},\mathbf{K}_{\epsilon_*|\mathbf{I}_\epsilon})
\end{equation}
where
\begin{equation}\label{eq:sol}
\begin{split}
    \boldsymbol\mu_{\bar{\epsilon}_*|\mathbf{I}_\epsilon} &= \mathbf{K}_*(\mathbf{K}_I+\sigma_n^2\mathbf{I})^{-1}\mathbf{I}_\epsilon \\
    \mathbf{K}_{\epsilon_*|\mathbf{I}_\epsilon} &= \mathbf{K}_{\epsilon}(\mathbf{x}_*,\mathbf{x}_*) - \mathbf{K}_*(\mathbf{K}_I+\sigma_n^2\mathbf{I})^{-1}\mathbf{K}_*^\Transp
\end{split}
\end{equation}
An analytic solution to the double integral is not known. However, computationally expensive numerical integration can be avoided by using an approximation scheme such as the one described in Section~\ref{sec:reducing_computation_complexity}. It is worth noting that the extension to non-convex geometry is straight forward \cite{jidling2018probabilistic}.


\section{Implementation} 
\label{sec:implementation}
Two practical aspects need to be considered when implementing the reconstruction algorithm; computational complexity, and hyperparameter selection. 

\subsection{Reducing Computation Complexity} 
\label{sec:reducing_computation_complexity}
The computational complexity of solving Equation~\eqref{eq:sol} is twofold: firstly, the construction of $\mathbf{K}_{II}$ requires the evaluation of a double integral for every unique pair of measurements; secondly, the time required to invert $\mathbf{K}_{II}+\sigma_n^2\mathbf{I}$ scales with $\mathcal{O}(N^3)$. 
X-ray strain tomography problems of the type considered here have a large number of measurements, motivating the use of an approximation scheme.
Here, we consider the approximation scheme proposed in \cite{solin2014hilbert} which has previously been used for strain estimation \cite{jidling2018probabilistic,hendriks2018robust}.
Using this scheme the covariance functions assigned to the Beltrami stress functions, $\mathbf{K}_{\Phi_i}$, can be approximated by a finite series of $m$ basis functions;
\begin{equation}
\begin{split}
    k_{i}(\mathbf{x},\mathbf{x}') &\approx \sum_{k=1}^{m} \phi_{i,k}(\mathbf{x})\Sigma_{pi,kk}\phi_{i,k}(\mathbf{x}') \\
    &= \boldsymbol\phi_{i}(\mathbf{x})\Sigma_{pi}\boldsymbol\phi_{i}(\mathbf{x}')^\Transp,
\end{split}
 \end{equation} 
where each $\phi_{i,k}(\mathbf{x})$ is a basis function and $\Sigma_{pi,kk}$ is its spectral density;
\begin{equation}
    \begin{split}
        \phi_{i,k}(x) &= \frac{1}{L_xL_yL_z}\sin(\lambda_{xk}(x+L_x))\\
        &\hspace{10mm}\sin(\lambda_{yk}(y+L_y))\sin(\lambda_{zk}(z+L_z)),\\
        \Sigma_{pi,k} & = \int K(\mathbf{r})\exp(\redchanges{\mathrm{-i}}\boldsymbol\lambda^\Transp\mathbf{r})\,\mathrm{d}r.
    \end{split}
\end{equation}
Here, $\mathbf{r} = \mathbf{x}-\mathbf{x}'$, and $\boldsymbol\lambda = [\lambda_{xj} \ \ \lambda_{yj} \ \ \lambda_{zj}]^\Transp$.
For the squared-exponential covariance function used in Section~\ref{sec:simulation} the spectral density is
\begin{equation}
    \Sigma_{pi,kk} = \sigma_f^2(2\pi)^\frac{3}{2}l_xl_yl_z\exp\left(-\frac{1}{2}\left(l_x^2\lambda_{xk}^2 + l_y^2\lambda_{yk}^2+l_z^2\lambda_{zk}^2\right)\right),\\ 
\end{equation}
where $\sigma_f$, $l_x$, $l_y$, and $l_z$ are commonly referred to as hyper-parameters and are discussed in Section~\ref{sec:hyperparameter_optimisation}. 
The parameters $L_x$, $L_y$, $L_z$, and $\boldsymbol\lambda$ are analogous to the frequency and phase of the basis functions.

In this work, they were chosen so that the basis functions spanned a region where their spectral densities were greater than a predefined minimum threshold. 
This helps to ensure numerical stability while capturing the dominant modes of the reconstruction.

By concatenating the basis functions we can concisely express the approximation for the covariance function $\mathbf{K}_\Phi$;
\begin{equation}
    \begin{split}
        \mathbf{K}_\Phi &\approx \boldsymbol\phi_\Psi \boldsymbol\Sigma_\Psi \boldsymbol\phi_\Psi^\Transp, \\ 
         \boldsymbol\phi_{\Psi} &= \pmat{\boldsymbol\phi_{1} \\ \boldsymbol\phi_{2} \\ \boldsymbol\phi_{3} \\ \boldsymbol\phi_{4} \\ \boldsymbol\phi_{5} \\ \boldsymbol\phi_{6}}^\Transp, \quad \boldsymbol\Sigma_\Psi = 
    \begingroup
    \setlength\arraycolsep{1.5pt}
    \pmat{\Sigma_{p1} & 0 & 0 & 0 & 0 & 0 \\
     0 & \Sigma_{p2} & 0 & 0 & 0 & 0 \\
      0 & 0 & \Sigma_{p3} & 0 & 0 & 0 \\
      0 & 0 & 0 & \Sigma_{p4} & 0 & 0 \\
      0 & 0 & 0 & 0 & \Sigma_{p5} & 0 \\
     0 & 0 & 0 & 0 & 0  & \Sigma_{p6} } \endgroup
     \end{split}
\end{equation}

It is now a straight forward application of the mappings \eqref{eq:beltrami_to_strain} and \eqref{eq:measurement_operator} to approximate the covariances required in Section~\ref{sec:strain_field_reconstruction} to reconstruct the strain field;

\begin{subequations}\label{eq:approximate_covariance_functions}
\begin{equation}
    \begin{split}
        \mathbf{K}_{\boldsymbol\epsilon} &\approx \boldsymbol\phi_{\epsilon}(\mathbf{x}_*)\boldsymbol\Sigma_\Psi\boldsymbol\phi_{\epsilon}(\mathbf{x}_*)^\Transp \\
        (\mathbf{K}_*)_i &\approx \boldsymbol\phi_{\epsilon}(\mathbf{x}_*)\boldsymbol\Sigma_\Psi\boldsymbol\phi_{\mathbf{I}}(\boldsymbol\eta_i)^\Transp \\
        (\mathbf{K}_\mathbf{I})_{ij} &\approx \boldsymbol\phi_{\mathbf{I}}(\boldsymbol\eta_i)\Sigma_\Psi\boldsymbol\phi_{\mathbf{I}}(\boldsymbol\eta_j)^\Transp
    \end{split}
\end{equation}  
where 
\begin{equation}
    \begin{split}
        \boldsymbol\phi_{\epsilon}(\mathbf{x}_*) &= \mathbf{H}\mathcal{B}^\mathbf{x}\boldsymbol\phi_{\Phi}(\mathbf{x}_*) = \boldsymbol\phi_{*} \\
        \boldsymbol\phi_{I,i}(\boldsymbol\eta_i) &= \mathcal{L}^\mathbf{x}(\boldsymbol\eta_i)\boldsymbol\phi_{\epsilon}(\mathbf{p}_i+\hat{\mathbf{n}}s) = \boldsymbol\phi_{I,i}  \hspace{10mm} i = 1,\dots,N\\
     \end{split}
\end{equation}
\end{subequations}
This simplifies the problem to the calculation of $\boldsymbol\phi_{\epsilon}(\mathbf{x}_*)$ and $\boldsymbol\phi_{I,i}(\boldsymbol\eta_i)$, which are intuitively the basis functions for the strain field and the measurements, respectively, and only require a single application of the mappings.
Complete expressions for these basis functions are found in Appendix~\ref{sec:approximation_basis_functions}.

The posterior mean and covariance are approximated by
\begin{equation}
\begin{split}
    \boldsymbol\mu_{\epsilon_*|\mathbf{I}} & \approx \boldsymbol\phi_*\left(\boldsymbol\phi_I^\Transp\sigma_n^{-2}\boldsymbol\phi_I + \Sigma_\Psi^{-1}\right)^{-1}\boldsymbol\phi_I^\Transp\sigma_n^{-2}\mathbf{I}_\epsilon\\
    \mathbf{K}_{\epsilon_* | \mathbf{I}}  &\approx \boldsymbol\phi_*\left(\boldsymbol\phi_I^\Transp\sigma_n^{-2}\boldsymbol\phi_I + \Sigma_\Psi^{-1}\right)^{-1}\boldsymbol\phi_*^\Transp
\end{split}
\end{equation}
This avoids forming the covariance matrices, reduces the complexity of the regression to $\mathcal{O}(Nm^2)$ and also removes the need for numerical derivatives.

\subsection{Hyperparameter Optimisation} 
\label{sec:hyperparameter_optimisation}
The covariance functions, assigned to the potential functions, are characterised by their hyperparameters $\boldsymbol\theta$. For example, the squared-exponential covariance function has hyperparameters $\boldsymbol\theta = \{\sigma_f,l_x,l_y,l_z\}$; where $\sigma_f$ encodes our prior uncertainty and the length scales $l_x$, $l_y$, and $l_z$ provide an assumption of smoothness. 
The hyperparmaters are selected by maximising the marginal log likelihood, $\log p(\mathbf{I}_\epsilon | \{\boldsymbol\eta_i\},\boldsymbol\theta)$;
\begin{equation}
\begin{split}
    \boldsymbol\theta_* &= \argmax_{\boldsymbol{\theta}}\Bigg[-\frac{1}{2}\log\det(\mathbf{K}_{II}+\sigma_n^2\mathbf{I}) \\ &\hspace{30mm}-\frac{1}{2}\mathbf{I}_\epsilon^\Transp(\mathbf{K}_{II}+\sigma_n^2\mathbf{I})^{-1}\mathbf{I}_\epsilon\Bigg],
\end{split}
\end{equation}
where $\mathbf{K}_{II}$ is a function of $\boldsymbol\theta$. For the approximation in Section~\ref{sec:reducing_computation_complexity} $\mathbf{K}_{II}$ is replaced by $\boldsymbol\phi_{I}\Sigma_p\boldsymbol\phi_I^\Transp$, and expressions for the approximate log likelihood and its derivatives can be found in \cite{hendriks2018robust} or \cite{jidling2018probabilistic}. The parameters can thereafter be optimised using a gradient-based method, such as the BFGS algorithm in \cite{wright1999numerical}.


\section{Simulation Demonstration} 
\label{sec:simulation}
\begin{figure}
    \centering
    \includegraphics[width=0.8\linewidth]{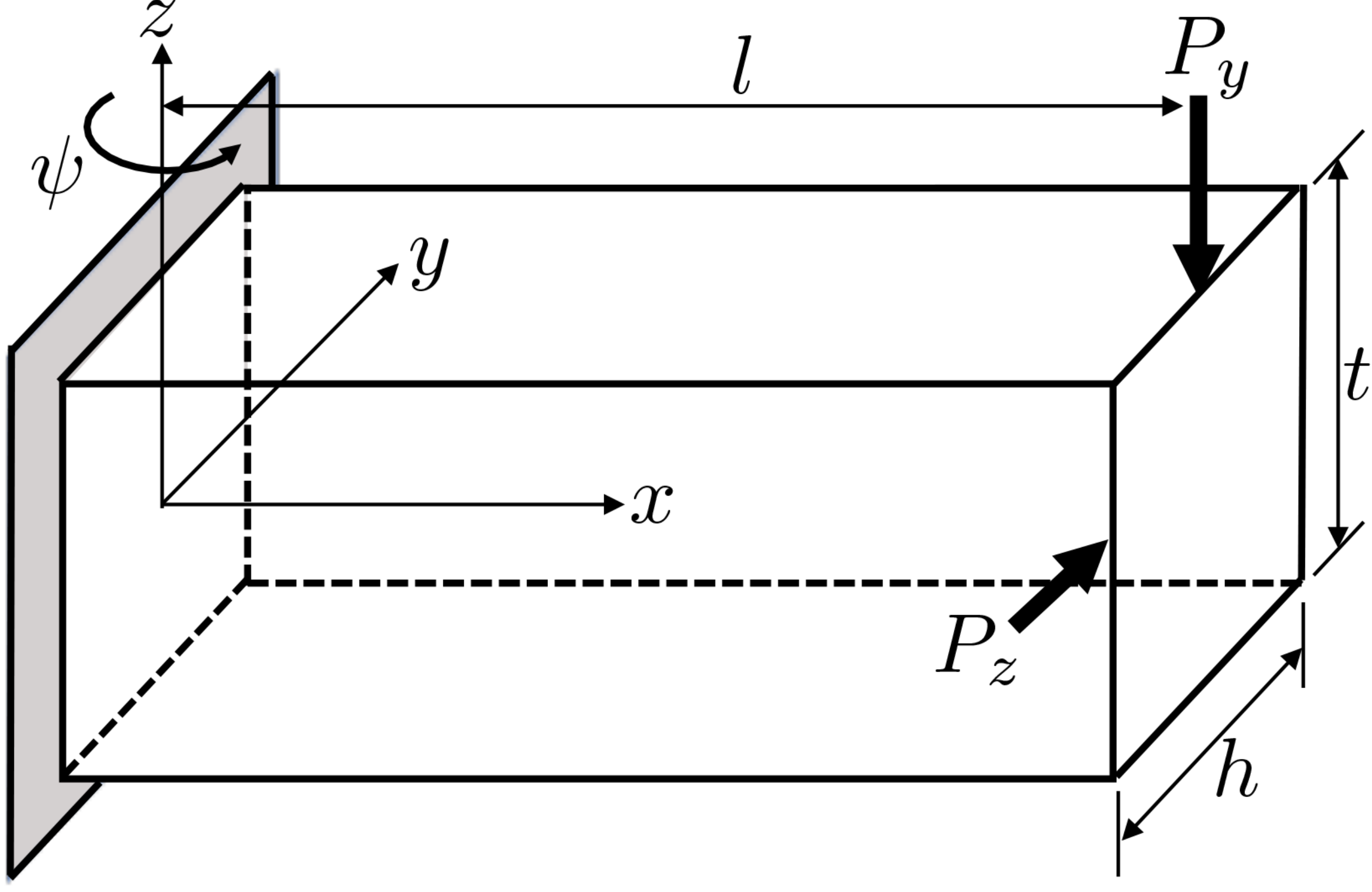}
    \caption{Three-dimensional Cantilever beam geometry and coordinate system with $l = 20\text{mm}$, $h=10\text{mm}$, $t=6\text{mm}$, $E = 200\text{GPa}$, $P_{y} = 2\text{KN}$, $P_{z} = 1\text{KN}$, $\nu = 0.28$, $I_{yy} = \frac{th^3}{12}$, and $I_{zz} = \frac{t^3h}{12}$. The triaxial strain field is given by a superposition of the Saint-Venant approximation to the strain field in the $xy$- and $yz$-planes as per Equation~\ref{eq:3D_CB_strain}. The sample is rotated in $\psi$ to give different projections of the strain field.}
    \label{fig:CB_geom}
\end{figure}

The method presented is demonstrated on a simulated 3D cantilevered beam example with the strain field given by a superposition of the Saint-Venant approximation to the strain field in the $xy$- and $yz$-planes.
This is chosen as an appropriate example in the absence of a suitable experimental data set.
The strain field is given by
\begin{equation}\label{eq:3D_CB_strain}
    \begin{split}
        \epsilon_{xx}(\mathbf{x}) &= \frac{P_y (l-x)y}{EI_{yy}}+\frac{P_z (l-x)z}{EI_{zz}} \\
        \epsilon_{yy}(\mathbf{x}) &= -\nu\frac{P_y (l-x)y}{EI_{yy}}-\nu\frac{P_z (l-x)z}{EI_{zz}} \\
        \epsilon_{zz}(\mathbf{x}) &= -\nu\frac{P_y (l-x)y}{EI_{yy}}-\nu\frac{P_z (l-x)z}{EI_{zz}} \\
        \epsilon_{xy}(\mathbf{x}) &= -(1+\nu)\frac{P_y(\frac{1}{4}h^2-y^2)}{EI_{yy}} \\
        \epsilon_{xz}(\mathbf{x}) &= -(1+\nu)\frac{P_z(\frac{1}{4}t^2-z^2)}{EI_{zz}} \\
        \epsilon_{yz}(\mathbf{x}) &= 0
    \end{split}
\end{equation}

where geometry and loading is defined in Figure~\ref{fig:CB_geom}.

Measurements were simulated through these strain fields using Equation \eqref{eq:measurement_operator}. 
The measurement geometry corresponded to rotating the sample about $z$ and for each angle using a $40\times40$ grid of incident beams.
\redchanges{A diffraction angle of $2\theta = 10^\circ$ was used for the simulation, giving the angle between the incident beam and the direction of measured strain as $\alpha=85^\circ$.}
For each incident beam 36 strain directions are measured, corresponding to using $10^\circ$ segments from the Debye-Scherrer rings.
The measurements were corrupted by zero-mean Gaussian noise with standard deviation $\sigma_m = 1\times10^{-4}$.

\begin{figure}[!ht]
    \centering
    \includegraphics[width=0.8\linewidth]{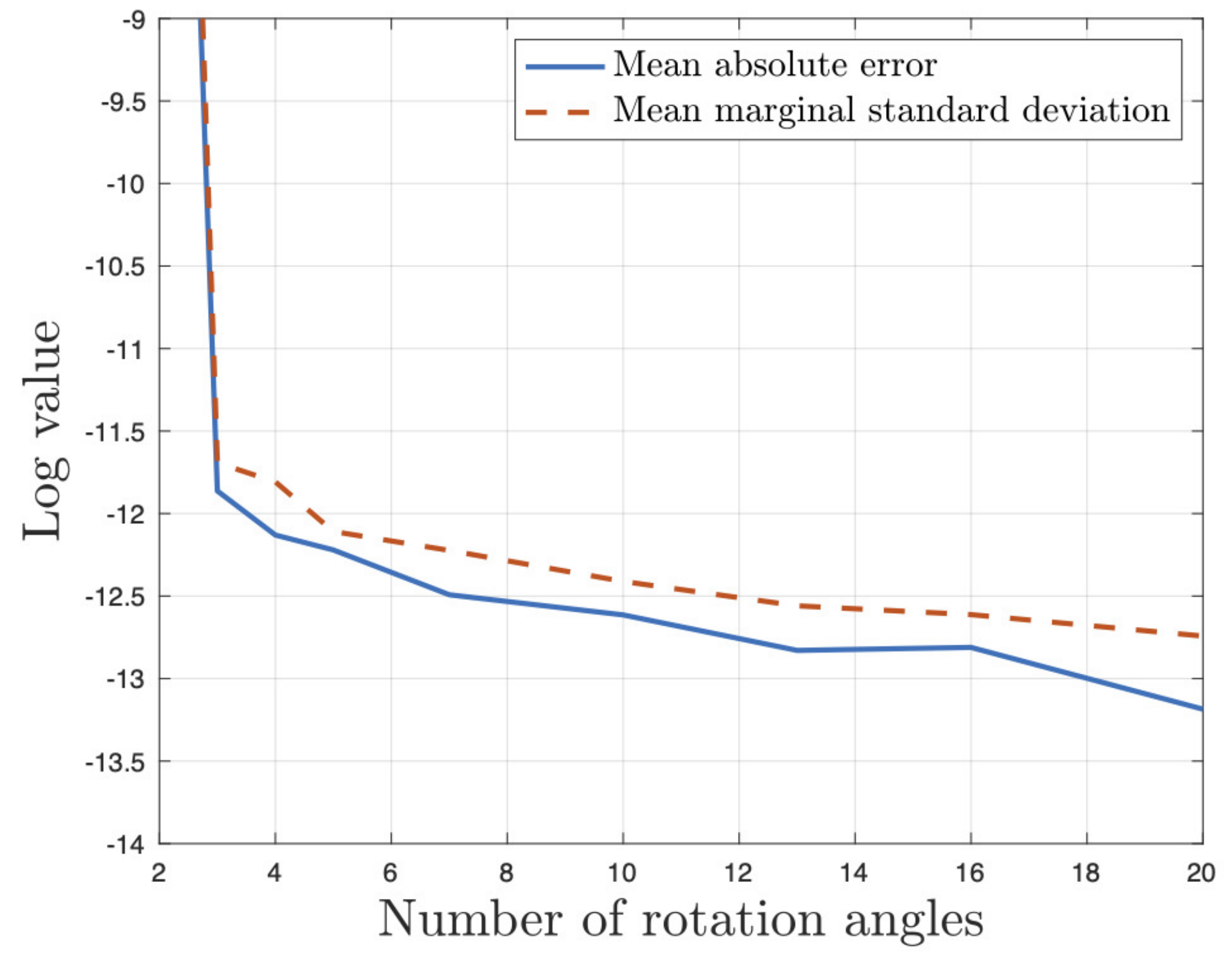}
    \caption{Convergence of the reconstruction in simulation as the number of rotation angles is increase. The logarithm of the mean absolute error is shown along with the logarithm of the mean marginal standard deviation of the predicted strains computed as $\text{mean}\left(\text{diag}(\mathbf{K}_{\epsilon_* | \mathbf{I}})^{1/2}\right)$.}
    \label{fig:conv_study}
\end{figure}

A convergence study as the number of rotation angles is increased was run and the results are shown in Figure~\ref{fig:conv_study}.
For each simulation the $n_\psi$ angles were chosen to be linearly spaced over $\frac{n_\psi}{n_\psi+1}180^\circ$.
From equation~\ref{eq:measurement_operator} it is clear that two beams with travelling along the same path but with opposite directions provide measurements of almost identical components of the strain field.
Therefore, it is not necessary to choose angles spanning $360^\circ$.
The mean absolute error and the mean marginal standard deviation of the predicted strains are reported. The strong correlation between the mean marginal standard deviation and the mean absolute error suggests that the posterior covariance $\mathbf{K}_{\epsilon_* | \mathbf{I}}$ could inform the user as to the expected error in the reconstruction.

These results indicate that at least three rotation angles are required for an accurate reconstruction. 
This is true regardless of the number of incident beams used per rotation angle, and intuitively can be explained as each rotation angle the incident beams predominantly provide information about the strains lying in the plane perpendicular to the beams.
Hence, three unique sets of these `in-plane' strains need to be observed to recover the three-dimensional strain field.

\begin{figure*}[!ht]
    \centering
    \includegraphics[width=0.9\linewidth]{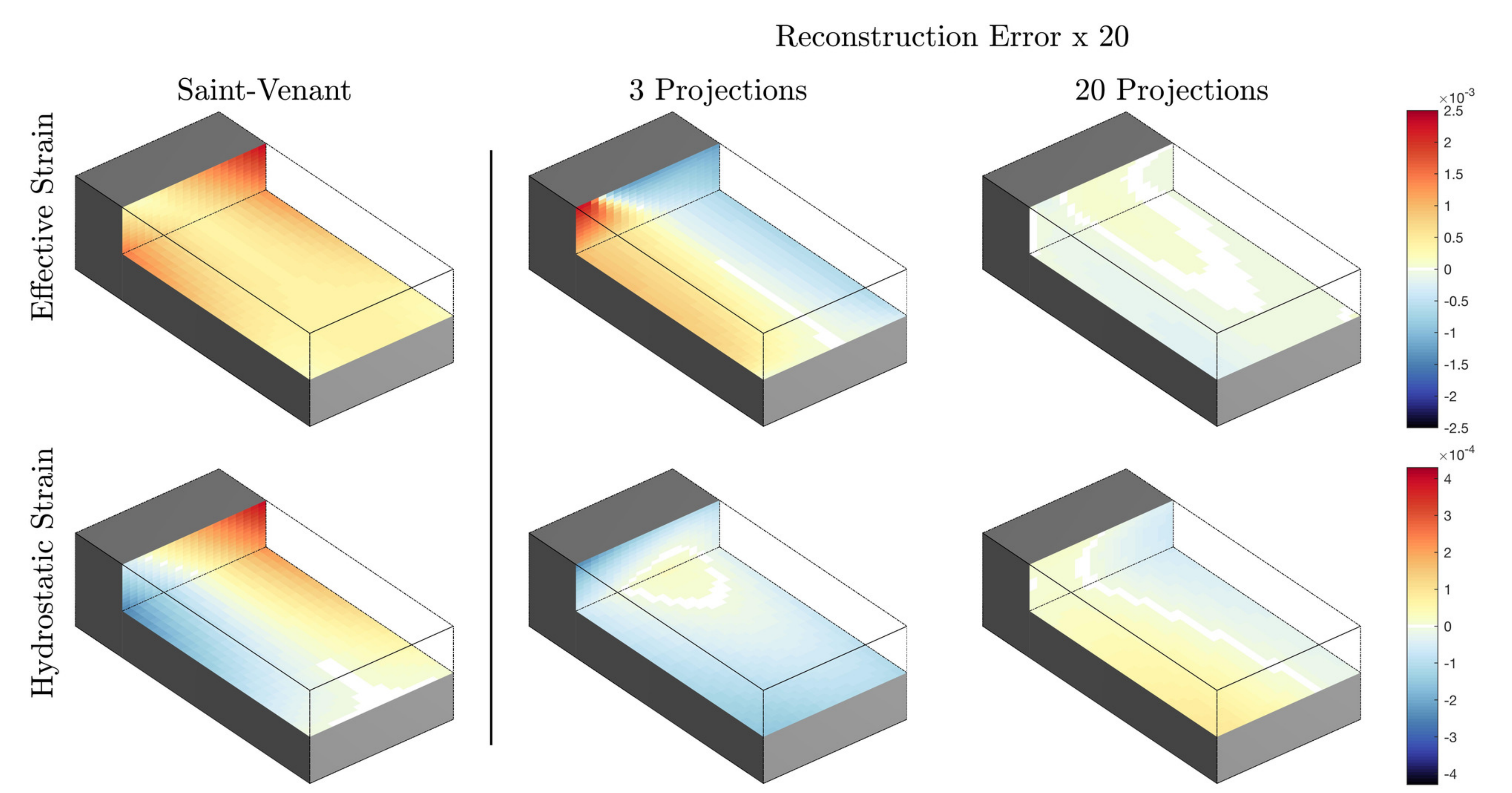}
    \caption{Shown in 3D is the reconstruction errors for 3 projections (centre) and 20 projections (right) alongside the Saint-Venant strain field (left).
    The effective strain is shown on the top row and the hydrostatic strain is shown on the bottom row.
    The magnitude of the error is reduced for 20 projections when compared to the results from 3 projections.
    The results are shown for two cutting planes; a plane at $z=0\text{mm}$, and a plane at $x=4\text{mm}$.
}
    \label{fig:result_eff_hyd}
\end{figure*}

The results of reconstructing from the measurement sets generated using three and twenty rotation angles are shown in Figure~\ref{fig:result_eff_hyd}. 
The results show the reconstructed effective and hydrostatic strains along two cutting planes, which give a good indication of the overall accuracy of the reconstruction.
The effective\footnote{The effective strain is the strain equivalent of von Mises stress sometimes referred to as the Mises effective strain.}, $\epsilon_{\text{eff}}$, and hydrostatic, $\epsilon_{\text{hyd}}$, strains are defined as
\begin{equation}
\begin{split}
    \epsilon_{\text{hyd}} &= \frac{1}{3}\left(\epsilon_{xx}+\epsilon_{yy}+\epsilon_{zz}\right) \\
    \epsilon_{\text{eff}} &= \Bigg(\frac{2}{3}\left(\epsilon_{xx}-\epsilon_{\text{hyd}}\right)^2+\frac{2}{3}\left(\epsilon_{yy}-\epsilon_{\text{hyd}}\right)^2+\\
    &\qquad \frac{2}{3}\left(\epsilon_{xx}-\epsilon_{\text{hyd}}\right)^2 + \frac{4}{3}\left(\epsilon_{xy}^2+\epsilon_{xz}^2+\epsilon_{yz}^2\right)\Bigg)^{1/2}
\end{split}
\end{equation}

\redchanges{Numerical simulations were also used to investigate the importance of $\alpha$ on the ability of the method to accurately reconstruct the the strain field.
Comparing the accuracy when the strain field was reconstructed from 10 projections with $\alpha$ ranging from 90 to 85 degrees it was found that the specific angle was not important for the proposed method to be able to accurately reconstruct the strain field.
For example, using $\alpha = 90^\circ$ gave a mean relative error of $1.04\%$ compared to $0.99\%$ for $\alpha = 85^\circ$.
This supports the argument that it is the assumption of equilibrium rather than measurement direction that are not quite perpendicular to the incident beam that allows the proposed method to successfully reconstruct the simulated strain field using single-axis tomography.}

\begin{figure}[tb]
    \centering
    \includegraphics[width=1.0\linewidth]{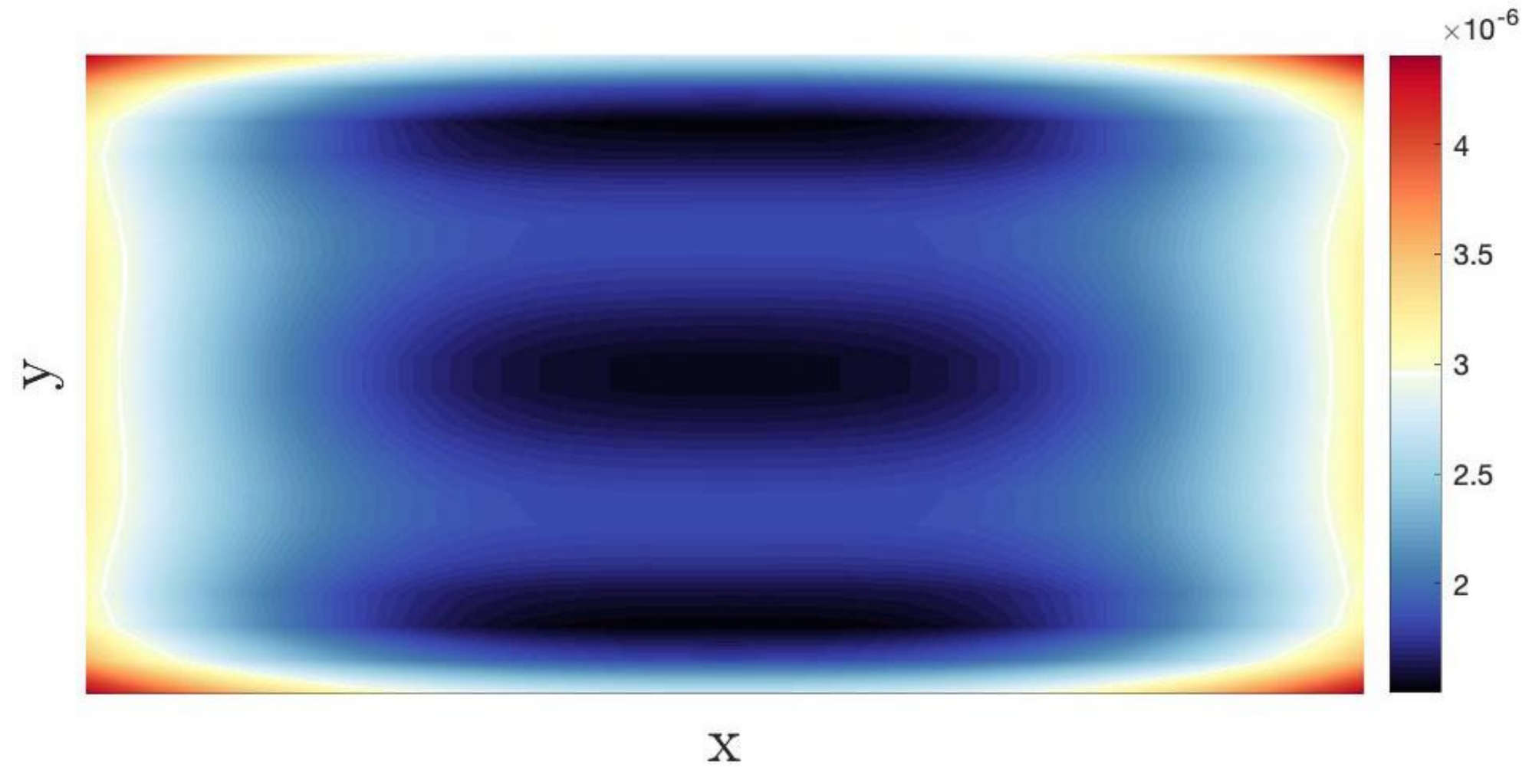}
    \caption{Standard deviation of the reconstructed $\epsilon_{xx}$ component from twenty simulated projections on the $z= 0\text{mm}$ cutting plane.}
    \label{fig:std_xx}
\end{figure}

In addition to the mean value for the reconstructed strain field, Equation~\ref{eq:sol} also provides a measure of the uncertainty. This can be used to determine where the reconstruction is least certain. For example, the marginal standard deviation of the reconstructed $\epsilon_{xx}$ component from $20$ projections is shown in Figure~\ref{fig:std_xx} for the cutting plane at $z= 0\text{mm}$. The standard deviation is highest near the sample boundary indicating that this is where the largest errors are likely to be. Standard deviation maps of the other strain components are similar.

\section{Conclusion} 
\label{sec:conclusion}
A Bayesian approach to triaxial strain tomography from high-energy X-ray diffraction measurements has been presented.
This approach models the strain field as a Gaussian process such that the resulting reconstruction always satisfies equilibrium.
From simulated measurements, this method was found to be capable of accurately reconstructing a full strain field in the presence of realistic levels of measurement noise. 
\redchanges{These numerical results suggest that it is possible to tomographically reconstruct a full triaxial strain distribution using single axis tomography by assuming equilibrium.}
While the squared-exponential covariance function proved suitable in this demonstration, other covariance functions, such as the Mat\'{e}rn or rational quadratic, may be more suitable for a particular strain field.

The next stage of this work will involve the planning and execution of an experimental demonstration of the technique. 
The ability to test the algorithm on simulated strain fields will provide guidance in this process.
For example, for a given strain field the trade-off between measurement noise, number of projection angles, and spacing of measurements for each projection can be investigated.


\section{Acknowledgements} 
\label{sec:acknowledgements}
This work is supported by the Australian Research Council through a Discovery Project Grant (DP170102324).


\appendix
\section{Approximation Basis Functions} 
\label{sec:approximation_basis_functions}
This section provides expressions for the basis functions given in \eqref{eq:approximate_covariance_functions} required for the approximation scheme used in Section~\ref{sec:reducing_computation_complexity}.

As stated earlier, a set of bases for the strain field is defined by a linear mapping from a set of bases for the Beltrami stress functions;
\begin{equation}
    \boldsymbol\phi_{\epsilon}(\mathbf{x}_*) = \mathbf{H}\mathcal{B}^\mathbf{x}\boldsymbol\phi_{\Phi}(\mathbf{x}_*).
\end{equation}
This is a linear combination of the partial derivatives of each Beltrami stress, i, function basis, k;
\small
\begin{equation}
\begin{split}
    \phi_{i,k} &= \frac{1}{\sqrt{L_xL_yL_z}}C_0, \\ 
    \frac{\partial^2}{\partial x^2}\phi_{i,k} &= \frac{-\lambda_{x,k}^2}{\sqrt{L_xL_yL_z}}C_0, \\
    \frac{\partial^2}{\partial y^2}\phi_{i,k} &= \frac{-\lambda_{y,k}^2}{\sqrt{L_xL_yL_z}}C_0, \\
    \frac{\partial^2}{\partial z^2}\phi_{i,k} &= \frac{-\lambda_{z,k}^2}{\sqrt{L_xL_yL_z}}C_0, \\
    \frac{\partial^2}{\partial x \partial y}\phi_{i,k} &= \frac{\lambda_{x,k}\lambda_{y,k}}{\sqrt{L_xL_yL_z}}C_1, \\
    \frac{\partial^2}{\partial x \partial z}\phi_{i,k} &= \frac{\lambda_{x,k}\lambda_{z,k}}{\sqrt{L_xL_yL_z}}C_2, \\
    \frac{\partial^2}{\partial y \partial z}\phi_{i,k} &= \frac{\lambda_{y,k}\lambda_{z,k}}{\sqrt{L_xL_yL_z}}C_3,
\end{split}
\end{equation}
\normalsize
where
\small
\begin{equation}
    \begin{split}
        C_0 = \sin\left(\lambda_{x,k}(x+L_x)\right)\sin\left(\lambda_{y,k}(y+L_y)\right)\sin\left(\lambda_{z,k}(z+L_z)\right), \\
        C_1 = \cos\left(\lambda_{x,k}(x+L_x)\right)\cos\left(\lambda_{y,k}(y+L_y)\right)\sin\left(\lambda_{z,k}(z+L_z)\right), \\
        C_2 = \cos\left(\lambda_{x,k}(x+L_x)\right)\sin\left(\lambda_{y,k}(y+L_y)\right)\cos\left(\lambda_{z,k}(z+L_z)\right), \\
        C_3 = \sin\left(\lambda_{x,k}(x+L_x)\right)\cos\left(\lambda_{y,k}(y+L_y)\right)\cos\left(\lambda_{z,k}(z+L_z)\right).
    \end{split}
\end{equation}
\normalsize

Applying the measurement mapping, $\mathcal{L}(\boldsymbol{eta}_i)$ to the basis functions for the strain field gives a set of basis functions for the $i^\text{th}$ measurement;
\begin{equation}
    \boldsymbol\phi_{I,i}(\boldsymbol\eta_i) = \mathcal{L}^\mathbf{x}(\boldsymbol\eta_i)\boldsymbol\phi_{\epsilon}(\mathbf{p}_i+\hat{\mathbf{n}}s).
\end{equation}
Which is a linear combination of the line integrals of the partial derivatives of each basis function. Defining
\small
\begin{equation}
\begin{split}
    \alpha_x &= \lambda_{x,k}(L_x+\hat{n}_1 s + x_0), \\
    \alpha_y &= \lambda_{y,k}(L_y+\hat{n}_2 s + y_0), \\
    \alpha_z &= \lambda_{z,k}(L_z+\hat{n}_3 s + z_0), \\
    \Gamma_1 &= \frac{\cos\left(\alpha_x - \alpha_y -\alpha_z\right) }{\lambda_{x,k}\hat{n}_1 - \lambda_{y,k}\hat{n}_2 - \lambda_{z,k}\hat{n}_3}, \\
    \Gamma_2 &= \frac{\cos\left(\alpha_x + \alpha_y -\alpha_z\right) }{\lambda_{x,k}\hat{n}_1 + \lambda_{y,k}\hat{n}_2 - \lambda_{z,k}\hat{n}_3}, \\
    \Gamma_3 &= \frac{\cos\left(\alpha_x - \alpha_y +\alpha_z\right) }{\lambda_{x,k}\hat{n}_1 - \lambda_{y,k}\hat{n}_2 + \lambda_{z,k}\hat{n}_3}, \\
    \Gamma_4 &= \frac{\cos\left(\alpha_x + \alpha_y +\alpha_z\right) }{\lambda_{x,k}\hat{n}_1 + \lambda_{y,k}\hat{n}_2 + \lambda_{z,k}\hat{n}_3}, \\
\end{split}
\end{equation}
\normalsize
the necessary components can be written as
\small
\begin{equation}
\begin{split}
    \int_0^L \frac{\partial^2\phi_{i,k}(\mathbf{p}_i+\hat{\mathbf{n}}s)}{\partial x^2}\,\mathrm{d}s &= \frac{-\lambda_{x,k}^2}{4\sqrt{L_xL_yL_z}}\left(\Gamma_1 - \Gamma_2 - \Gamma_3 + \Gamma_4\right), \\
    \int_0^L \frac{\partial^2\phi_{i,k}(\mathbf{p}_i+\hat{\mathbf{n}}s)}{\partial y^2}\,\mathrm{d}s &= \frac{-\lambda_{y,k}^2}{4\sqrt{L_xL_yL_z}}\left(\Gamma_1 - \Gamma_2 - \Gamma_3 + \Gamma_4\right), \\
    \int_0^L \frac{\partial^2\phi_{i,k}(\mathbf{p}_i+\hat{\mathbf{n}}s)}{\partial z^2}\,\mathrm{d}s &= \frac{-\lambda_{z,k}^2}{4\sqrt{L_xL_yL_z}}\left(\Gamma_1 - \Gamma_2 - \Gamma_3 + \Gamma_4\right), \\
    \int_0^L \frac{\partial^2\phi_{i,k}(\mathbf{p}_i+\hat{\mathbf{n}}s)}{\partial x \partial y}\,\mathrm{d}s &= \frac{\lambda_{x,k}\lambda_{y,k}}{4\sqrt{L_xL_yL_z}}\left(\Gamma_1 + \Gamma_2 - \Gamma_3 - \Gamma_4\right), \\
    \int_0^L \frac{\partial^2\phi_{i,k}(\mathbf{p}_i+\hat{\mathbf{n}}s)}{\partial x \partial z}\,\mathrm{d}s &= \frac{\lambda_{x,k}\lambda_{z,k}}{4\sqrt{L_xL_yL_z}}\left(\Gamma_1 - \Gamma_2 + \Gamma_3 - \Gamma_4\right), \\
    \int_0^L \frac{\partial^2\phi_{i,k}(\mathbf{p}_i+\hat{\mathbf{n}}s)}{\partial y \partial z}\,\mathrm{d}s &= \frac{\lambda_{y,k}\lambda_{z,k}}{4\sqrt{L_xL_yL_z}}\left(-\Gamma_1 - \Gamma_2 - \Gamma_3 - \Gamma_4\right), \\
\end{split}
\end{equation}




\bibliography{References}

\end{document}